\documentclass[preprint,12pt]{elsarticle}

\usepackage{amssymb}
\usepackage{amsmath}

\journal{Nuclear Physics B}

\begin{document}

\begin{frontmatter}

\title{Megastructures of type-III  civilizations orbiting galaxies}

\author{Zaza N. Osmanov} 

\affiliation{organization={School of Physics, Free University of Tbilisi}, 
            city={Tbilisi},
            postcode={0183}, 
            country={Georgia}}

\affiliation{organization={E. Kharadze Georgian National Astrophysical Observatory}, 
            city={Abastumani},
            postcode={0301}, 
            country={Georgia}}

\begin{abstract}
The article discusses the possibility of a Type-III extraterrestrial civilization constructing megastructures around a galaxy in regions where the galactic radiation becomes indistinguishable from the CMB radiation. For a Milky Way–like galaxy, we estimated the corresponding distance from its center at which a solar mass megastructure would need to be placed. We also showed that, from an energetic standpoint, placing such a massive object into orbit poses no fundamental difficulties. The detectability of the megastructure was also addressed.

\end{abstract}

\begin{keyword}
SETI, technosignatures, Dyson spheres
\end{keyword}

\end{frontmatter}

\section{Introduction}
\label{sec1}

In recent years, debates in the scientific literature about the detection of possible extraterrestrial civilizations have intensified. However, the search for a radio message is an important project, yet it is easy to show that even if several thousand technological civilizations exist in our Galaxy, communication would be stretched unrealistically over time. Therefore, in the last century, when SETI science was taking its first steps, Freeman Dyson proposed a highly original hypothesis: he suggested that if a so-called Type-II techno-civilization exists (according to the Kardashev classification \cite{kardashev}) — that is, a civilization that uses the entire energy output of its host star — then it would need to enclose the star within a large spherical shell (Dyson sphere). As a result, this spherical shell would heat up and appear in the sky as a gigantic infrared technosignature \cite{dyson}. It is worth noting that there has been some progress in the search for such objects, and several Dyson sphere candidates have already been identified \cite{carrigan,zackrisson}.  

\begin{figure}
  \centering {\includegraphics[width=11cm]{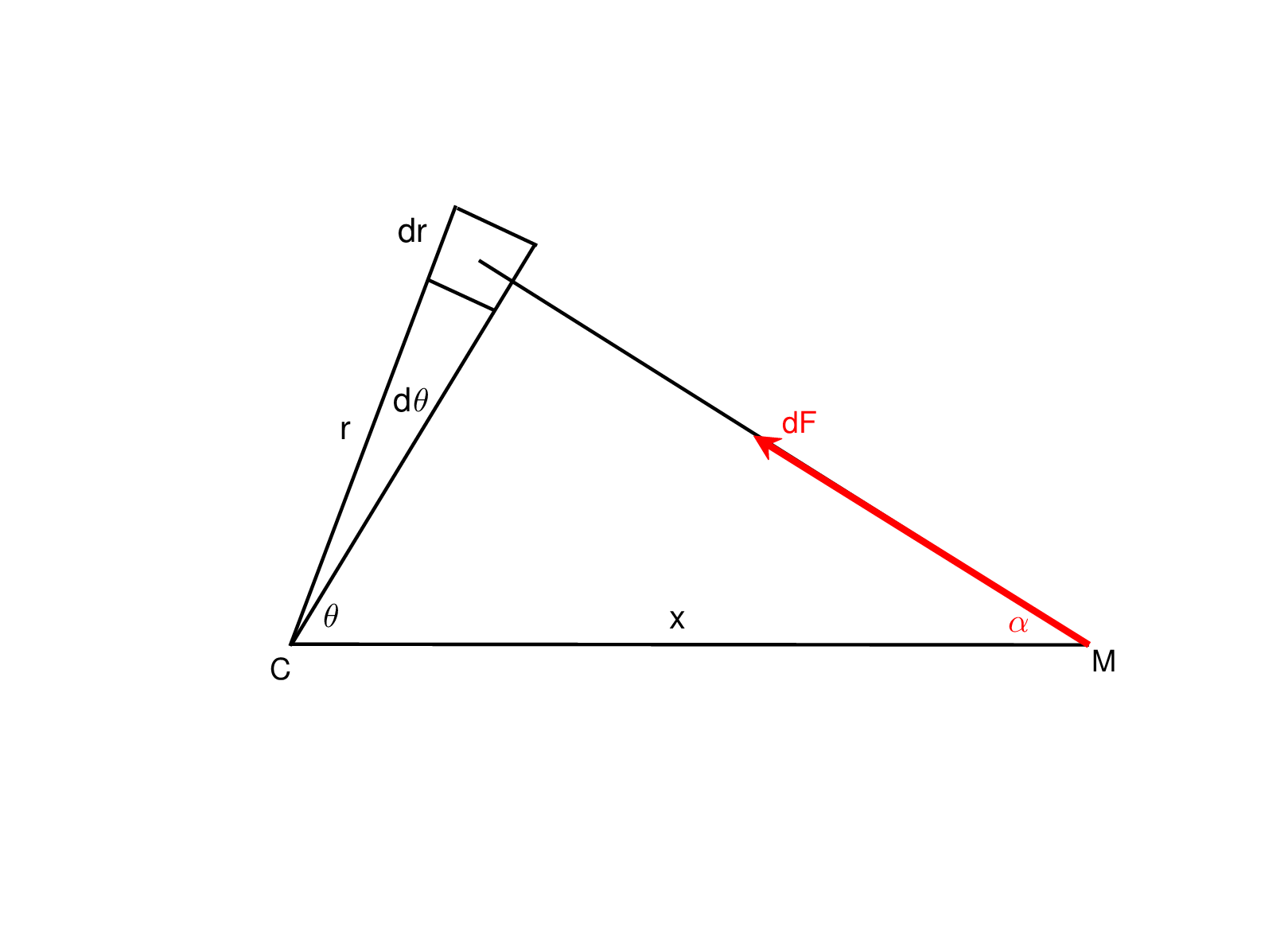}}
  \caption{A schematic illustration showing the locations of the galactic center (C) and the megastructure (M). We also show the distances and the differential force ${\bf dF}$, which represents the gravitational force exerted on the megastructure by an infinitesimal region located in the plane of the galaxy.}\label{fig1}
\end{figure}

Several extensions of Dyson’s idea have been proposed; in particular, authors have considered the construction of such megastructures around white dwarfs \cite{WD} and black holes \cite{BH}, while for pulsars a ring-type megastructure has been suggested \cite{puls1,puls2,haliki}. The interest in megastructures is so broad that researchers are even considering whether megastructures created by our own civilization could be detectable by extraterrestrial civilizations \cite{acta}.

Over the past decade, interest in technosignature research has grown to the extent that several interesting studies have been dedicated to megastructures built by Type-III civilizations (harnessing the energy of an entire galaxy) \cite{civ1,civ2,civ3}.

In the present paper, we examine the possibility of a megastructure orbiting a galaxy relatively far from its center built by Type-III extraterrestrial civilizations.

The paper is organized as follows: In section 2,  we examine the dynamics of the megastructure and estimate the required kinematic parameters for the Milky Way, and in the Summary we briefly discuss the obtained results.

\section{Discussion and Results}

In the literature, there is an active discussion of megastructures that might be constructed by highly advanced extraterrestrial civilizations in the course of their cosmic expansion \cite{cirkovic}. The reason may be the desire for space exploration, conducting experiments for scientific research, using energy, or other purposes. Whatever the reason may be, it is undeniable that such megastructures would inevitably leave traces in the sky, which might be, in principle, detectable. 

When searching for messages sent by extraterrestrial civilizations, the widely accepted view is that we should first look for signals based on universal knowledge, facts about the cosmos that any intelligent being inhabiting the universe would necessarily be aware of. For example, Cocconi and Morrison, in their well-known paper \cite{cocconi}, emphasize that since hydrogen is the most abundant element in the universe, it is reasonable to expect that a message might be sought at the frequency of neutral hydrogen emission.

It is clear that near the Galaxy its radiation is quite strong and gradually decreases with increasing distance. If a civilization, for certain reasons, wishes to build a "laboratory" in a region where the radiation is minimal, this makes sense only from a certain distance, because the cosmic microwave background (CMB) is present everywhere \cite{carroll}. Therefore, the corresponding minimum distance must satisfy the condition that the total Galactic radiation flux and the CMB background radiation flux have to be of the same orders of magnitude. Assuming the exponential distribution of luminosity density in the Milky Way galaxy, for the galactic intensity one obtains the following expression (see Fig. 1): 
\begin{equation}
\label{GFlux1} 
I_G\simeq\frac{\Sigma_{0}}{4\pi}\int_{0}^{\infty}\int_{0}^{2\pi}\frac{re^{-r/r_d}drd\theta}{x^2+r^2-2rx \cos{\theta}},
\end{equation}
where $\Sigma_{0}$ and $r_d$ are respectively the central luminosity density of and the scale length of the galaxy \cite{LG}.
By tedious but straightforward calculations, it is straightforward to show that the above integral approximately takes the form:
\begin{equation}
\label{GFlux2} 
I_G\simeq\frac{\Sigma_{0}}{4}\times\frac{r_d}{x},
\end{equation}
If we assume $I_G\lesssim\sigma T_{_{CMB}}^4$ where $\sigma$ denotes the Stefan-Boltzmann constant and $T_{_{CMB}}\simeq 2.7$ K \cite{carroll}, one can show 
\begin{equation}
\label{x1} 
\frac{x}{R}\gtrsim\frac{\Sigma_{0}r_d}{4\sigma T_{_{CMB}}^4R},
\end{equation}
where $x$ is the distance from the megastructure to the galactic center (see Fig. 1), and $R$ denotes a radius of the galaxy. Normalization by Milky Way parameters leads to
\begin{equation}
\label{x2} 
\frac{x}{R}\gtrsim 5.5\times\frac{\Sigma_{0}}{\Sigma_{_{MW}}}\times\frac{r_d}{r_{_{MW}}}\times\frac{R{_{_{MW}}}}{R}.
\end{equation}
where $\Sigma_{_{MW}}\simeq 900L_{\odot}/pc^2$, and $r_{_{MW}}\simeq 2.5$ kpc are respectively the central luminosity density of the Milky Way and its scale length \cite{LG}, $L_{\odot}\simeq 3.8\times 10^{33}$ erg s$^{-1}$ denotes the solar luminosity \cite{carroll} and $R{_{_{MW}}}\simeq 13.5$ kpc represents the radius of the Milky Way \cite{carroll}. From Eq. (\ref{x2}) it is clear that $x/R>>1$ and therefore, to estimate the orbiting speed, as a first approximation, we can avoid integrating the force ${\bf dF}$ and instead use an approximate expression $F\simeq GMm/x^2$, where $G$ denotes the gravitational constant and $M$ and $m$ represent the masses of the galaxy and the megastructure, respectively. Then, for the orbital velocity, one obtains
\begin{equation}
\label{vel} 
\upsilon\simeq\left(\frac{GM}{x}\right)^{1/2}\simeq 270\times\left( \frac{\Sigma_{0}}{\Sigma_{_{MW}}}\times\frac{r_d}{r_{_{MW}}}\times\frac{R{_{_{MW}}}}{R}\times\frac{M}{M{_{_{MW}}}}\right)^{-1/2}\; km\; s^{-1},
\end{equation}
where $M{_{_{MW}}}\simeq 1.26\times 10^{12}\; M_{\odot}$ \cite{Mg} and $M_{\odot}\simeq 2\times 10^{33}$ g is the solar mass. It is worth noting that the velocities of the probes Voyager 1 and 2, launched by NASA in the $1970$s, are respectively about $17$ km/s (Voyager-1) and $15$ km/s (Voyager-2)
\footnote{https://science.nasa.gov/mission/voyager/where-are-voyager-1-and-voyager-2-now/}. Therefore, it should not be surprising if a highly advanced civilization is capable of achieving velocities an order of magnitude higher.

One can straightforwardly check that reaching such velocities should not be a problem from an energetic point of view. In particular, $GM/x$ is the energy scale for a unit-mass body to lift it to the mentioned orbit and set it into rotation. For the Milky Way galaxy the energy scale is of the order of $\phi\simeq 7\times 10^{14}$ erg. Even if one assumes the journey with the speed of light, the travel time, $t$, becomes of the order of $10^5$ yrs. If we take into account that our current rate of energy consumption is $P_0\simeq 1.5\times 10^{20}$ erg s$^{-1}$ \cite{spaceX}, and assume a $1\%$ ($\eta = 0.01$) of annual technological growth, it turns out that to reach a Type-III level, a civilization at our current technological level would require 
\begin{equation}
\label{tau} 
\tau = \frac{ln(L_{_{MW}}/P_0)}{ln(1+\eta)}\simeq 6\times 10^3 yrs,
\end{equation}
where $L_{_{MW}}\simeq 10^{44}$ erg s$^{-1}$ is the Milky Way's total luminosity. It is evident that $\tau<<t$, indicating that any civilization that attempts, for any purpose, to build a megastructure away from its galaxy must belong to a highly advanced civilization, most likely a Type-III civilization. It is easy to estimate that transporting a solar-mass megastructure to the distance, $x$, requires an energy of the order of $GMM_{\odot}/x\simeq 1.5\times 10^{48}$ erg, which is equivalent to the energy that a Milky Way-like galaxy would radiate just over a couple of hours. Therefore, for a technologically super-advanced extraterrestrial civilization of level-III, constructing such a megastructure in the mentioned orbit should not be a problem.

The presence of an object in an orbit at a distance of about $5.5 R$ or greater may indicate its artificial origin, since this value (for a different galaxy the value of $x$ will be different) is the location beyond which the intensity of galactic radiation no longer exceeds the intensity of microwave radiation, and may point to an artificial origin of choice of the location. Therefore, one of the key objectives in the search for technosignatures should be to look for megastructures in such orbits.

If such objects are part of the CMB background in the context of radiation, then how can they be detected? Measuring the anisotropy of the microwave radiation could in principle allow this, but the resolution is so small that with present-day instruments this is practically impossible. On the other hand, if the megastructure is sufficiently massive, it may contribute to microlensing. In particular, the angular resolution of the Einstein ring caused by a megastructure $\theta\approx\sqrt{GMD/(4c^2)}$ \cite{lensing} on a distance $D = 1$ Mpc equals $\sim 90$ micro-arcsecond, which looks sufficiently promising \cite{gaia}. However, we plan to discuss the detection issue in detail in a forthcoming publication, and at this stage we only aimed to demonstrate that the detection of such megastructures does not pose a fundamental problem.

\section{Conclusion}

We considered the construction of megastructures, using the Milky Way as an example, on orbits where the galaxy’s radiation no longer exceeds the intensity of the microwave background radiation. 

We showed that the radius of such an orbit exceeds the galaxy’s radius by several times. 

It was also demonstrated that, from an energetic point of view, even placing a solar-mass object into such an orbit does not pose an energy problem for Type-III civilizations.

We also discussed the possibility of detecting these objects through microlensing and evaluated the angular resolution, which appears promising in the context of current instruments.

\section{Declaration of competing interest}
The authors report no conflict of interest.

\section{Acknowledgments}
The research was supported by the Shota Rustaveli National Science Foundation of Georgia (SRNSFG) Grant: FR-24-1751.

\end{document}